\documentclass[%
 reprint,
 amsmath,amssymb,
 aps,
 prx
]{revtex4-2}
\usepackage{tabularx}
\newcolumntype{Y}{>{\centering\arraybackslash}X}

\usepackage{graphicx}% Include figure files
\usepackage{dcolumn}% Align table columns on decimal point
\usepackage{bm}% bold math
\begin{document}

\title{Modelling the movements of organisms by stochastic theory in a comoving frame}

\author{Norberto Lucero-Azuara}
\email{n.lucero-azuara@qmul.ac.uk}
\affiliation{Centre for Complex Systems, School of Mathematical
  Sciences, Queen Mary University of London, Mile End Road, London E1
  4NS, United Kingdom}

\author{Rainer Klages}
\email{r.klages@qmul.ac.uk}
\affiliation{Centre for Complex Systems, School of Mathematical
  Sciences, Queen Mary University of London, Mile End Road, London E1
  4NS, United Kingdom}
\affiliation{London Mathematical Laboratory, 8 Margravine Gardens,
  London W6 8RH, United Kingdom}

\date{\today}

\begin{abstract}
Imagine you walk in a plane. You move by making a step of a certain
length per time interval in a chosen direction. Repeating this process
by randomly sampling step length and turning angle defines a
two-dimensional random walk in what we call {\em comoving frame}
coordinates. This is precisely how Ross and Pearson proposed to model
the movements of organisms more than a century ago. Decades later
their concept was generalised by including persistence leading to a
correlated random walk, which became a popular model in Movement
Ecology. In contrast, Langevin equations describing cell migration and
used in active matter theory are typically formulated by position and
velocity in a fixed Cartesian frame. In this article, we explore the
transformation of stochastic Langevin dynamics from the Cartesian into
the comoving frame. We show that the Ornstein-Uhlenbeck process for
the Cartesian velocity of a walker can be transformed exactly into a
stochastic process that is defined {\em self-consistently} in the
comoving frame, thereby profoundly generalising correlated random walk
models. This approach yields a general conceptual framework how to
transform stochastic processes from the Cartesian into the comoving
frame. Our theory paves the way to derive, invent and explore novel
stochastic processes in the comoving frame for modelling the movements
of organisms. It can also be applied to design novel stochastic
dynamics for autonomously moving robots and drones.
%or steering autonomously moving robots and drones.
\end{abstract}

\maketitle

%\tableofcontents

\section{\label{sec:level1}Introduction}

Organisms living at very different spatio-temporal scales, from moving
in the microworld to foraging across the surface of the earth, display
highly complex, random-looking migration patterns \cite{KLK26}. By now
there exists a wealth of experimental recordings of these patterns for
a huge variety of species, from insects \cite{KaSh83,LCK13,Wehn20} to
fishes \cite{Sims08,Sims10,BPA23}, birds \cite{Vis96,Edw07,VCTBN25},
mammals \cite{RMM04,RCM22,MCS23} and even humans
\cite{BHG06,GHB08,RWGMMP14}. Novel biologging techniques developed
over the past decade are delivering bigger and more precise data sets
\cite{Kays15,Nath22,WaPa23}. These developments pose the fundamental
challenge to understand the experimentally recorded organismic
movement patterns by constructing mathematical models from data
\cite{GiMa26},

Around 1905 Ross and Pearson introduced random walks for modelling the
migration of organisms, described in terms of step length and turning
angle with respect to the previous step for movements in a plane
\cite{Ross04,Ross05,Pea06}. In the simplest case, one may choose a
constant step length while the turning angle is sampled independently
and identically distributed from a uniform probability distribution,
reflecting the randomness of organismic movement \cite{Pea05,Pea06}.
This simple model became popular as the drunkard's walk
\cite{Pea06,Reif}, since its dynamics does not contain any memory,
which mathematically corresponds to a Markov process
\cite{vK,Gard09}. It is at the heart of describing diffusive
spreading in nature, technology and society \cite{BCKV18}. For decades
simple random walks have been used to model organismic movements
\cite{Ske51} until in the 1980's they were generalised by sampling the
turning angle from a unimodal distribution, which imposes a
correlation between two subsequent steps modelling one-step
persistence in organismic motion \cite{KaSh83}. In addition, one may
choose the step length as an independent and identically distributed
random variable. This model became known as a correlated random walk
(CRW) \cite{CPB08}. Together with state space, \cite{PPL17,ANC21},
hidden Markov \cite{PPL17,GAL23} and other stochastic models
\cite{MCB14,CPB08} CRWs form the theoretical backbone of {\em Movement
  Ecology} (ME), a field founded in 2008 \cite{Nath08,JPB18} mainly
driven by experimental biologists \cite{Nath22,Kays15}, which
endeavours to understand the movements of organisms, especially on
larger scales, in view of their interactions with the environment. One
may ask, however, whether including one-step persistence is sufficient
to fully understand the movements of organisms, which often feature
long-term memory way beyond a single step.

Another fundamental approach to model movement patterns draws on the
observation that they may look similar to the Brownian motion of a
tracer particle in a fluid, described by the famous Langevin equation
(LE) \cite{Lang08}. This equation is based on Newton's second law by
decomposing the force acting onto a Brownian particle into Stokes
friction and random collisions with the surrounding molecules,
modelled by Gaussian white noise. Formulated in terms of Cartesian
position and velocity of a moving particle, it is sometimes called
Newton's law of stochastic physics.  Mathematically, the LE represents
a Markovian Ornstein-Uhlenbeck (OU) process \cite{OU1930} for the
velocities of the Brownian particle \cite{CKW04}. Like simple random
walks, OU processes have been, and still are, widely used to model
organismic movement \cite{FSC15,PPL17}. In parallel, however, Langevin
dynamics was generalised to reproduce the non-trivial dynamics of
migrating cells \cite{NHKD17,MSDRN20,Diet22,KMBSN24,Kla24} and active
particles \cite{Schw03,RBELS12,BeDiL16,Schw19}.

\vspace*{-1.5cm}
\begin{figure}[hbt]
\centering
\includegraphics[width=1\columnwidth]{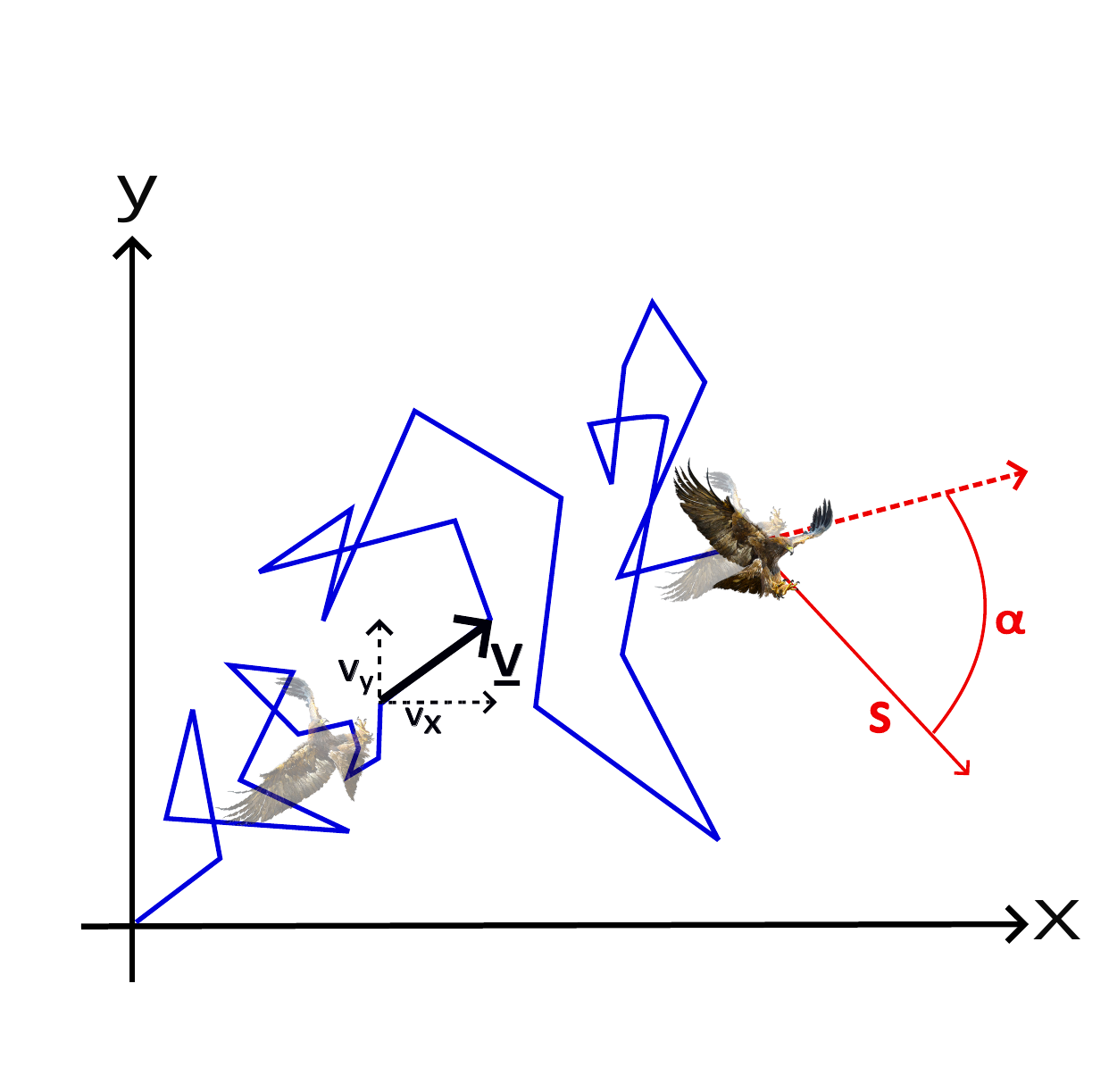}

\vspace*{-1.2cm}
\caption{Time-discrete trajectory (blue) of an organism (the pictures
  show an eagle) in a fixed Cartesian frame (black horizontal and
  vertical lines). Also shown is the velocity of the eagle at some
  point along the trajectory in Cartesian coordinates $v_x,v_y$. In
  contrast, at the final point of the trajectory we represent the
  movement of the eagle in comoving frame coordinates (red) in terms
  of speed $s$ and turning angle $\alpha$, where the latter is defined
  as the angle between subsequent velocity directions (dashed,
  respectively bold red lines). }\label{fig:eagle}
\end{figure}

Crucially, CRWs and the (generalised) LEs referred to above are
formulated in very different coordinate frames. What we call {\em
  comoving frame} coordinates traces back to Ross and Pearson's
original formulation of two-dimensional random walks, for which they
used step length and turning angle, as illustrated in
Fig.~\ref{fig:eagle}. Note that in this figure the step length is
replaced by the speed of an organism, where the speed is defined as
step length per time interval. In that sense, with comoving frame we
denote a frame attached to the center of mass of an organism
cotranslating and -rotating with its movements, whose abscissa is
aligned with the velocity of an organism, and the rotations of this
velocity vector yield the associated turning angle. Comoving frame
coordinates are biologically very well motivated, as it seems natural
to think of higher-dimensional movements of organisms in terms of step
length and turning angle \cite{Ross04,Pea05,Pea06}. Correspondingly,
step length distributions are intimately related to step selection
functions and associated concepts widely applied in ecology and
conservation to experimentally characterize the movements of organisms
\cite{TCB14,SRMVVBL20,Gunner24}. LEs, in contrast, are typically
defined in terms of Cartesian position and velocity of a moving
particle \cite{Lang08,vK,Gard09,Reif}. This description makes perfect
sense for representing the passive motion of a tracer particle driven
by collisions from the surrounding molecules in a fluid that as a
whole is at rest in a fixed Cartesian frame. It also simplifies the
theoretical analysis of these equations, especially if they are
generalised by including memory kernels
\cite{NHKD17,MSDRN20,Diet22,KMBSN24,Kla24}. However, one may
fundamentally question whether the Cartesian approach is correct for
modelling self-propelled, active movement that is generated
intrinsically by an organism itself, instead of having a passive
particle solely driven extrinsically by interactions with the
environment \cite{KLK26}. One may argue that active fluctuations
emerging internally within an organism should rather be modelled by a
stochastic process defined {\em self-consistently} in a frame comoving
with this organism, i.e., without explicitly involving any other
coordinates than speed and turning angle, and not by noise or friction
terms somewhat formulated externally with respect to a fixed Cartesian
frame. This problem is indeed taken into account, to some extent, in
some active particle models \cite{RoSch11,GSR12,RBELS12,BeDiL16},
where upon closer scrutiny one popular type of them
\cite{HJR07,BeDiL16} turns out to be identical to the CRW of ME
\cite{KLK26}. However, as we will demonstrate in this article, to
fully solve this problem one has to go one significant step
further. There are thus many reasons to investigate merging the
approaches of CRWs in a comoving frame and LEs in a fixed Cartesian
frame for constructing a more general organismic movement model that
combines advantages from both theories.

For constructing a stochastic model of bumblebee flights from
experimental data, in 2013 Lenz et al.\ addressed this problem by
fusing \textit{ad hoc} generalised LEs with CRWs in terms of two
coupled stochastic differential equations for speed and turning angle,
both of Langevin type but with {\em per se} arbitrary friction and
noise terms \cite{LCK13}. All these terms were extracted from
experimental data. Interestingly, the turning angle distribution was
found to be unimodal, as in a CRW, all noise terms were correlated,
and the relevant friction term was speed-dependent, as for a specific
active particle model \cite{Schw03,RBELS12}. Very recently, the same
approach of formulating generalised Langevin dynamics in a comoving
frame was exploited for experimentally steering superparamagnetic
colloidal microrobots with tailored statistics generating non-Brownian
anomalous diffusion, like fractional Brownian motion, by fine-tuning
magnetic fields \cite{GKV25}. The underlying general framework thus
promises to cross-link \cite{KLK26} the different big fields of ME
\cite{Nath08,JPB18,GiMa26}, Active Particles
\cite{Schw03,RBELS12,BeDiL16,Schw19} and Anomalous Diffusion
\cite{MeKl00,CKW04,MJCB14,KRS08,ZDK15} for modelling and understanding
the movements of organisms. This motivates to explore how to
analytically derive the equations stipulated in \cite{LCK13} from
first principles.

Our article solves this problem for the paradigmatic example of the OU
process defined in the Cartesian frame.  In Sec.~\ref{sec:level2} we
first show that Cartesian and comoving coordinates are cross-linked by
polar coordinates. We thus have to carefully distinguish between three
different frames for our transformation of stochastic processes, which
are the Cartesian, the polar and the comoving frame. As a 
tutorial warm-up exercise, in Sec.~\ref{sec-RW} we illustrate the
transformation between these three different frames for the example of
a simple Markovian random walk defined in the Cartesian frame by
deriving corresponding stochastic models in the other two
frames. Analytical results for all three models are compared with
computer simulations. In Sec.~\ref{sec-ou-comoving} we apply the same
conceptual framework to analytically construct different models in
the polar and in the comoving frame for the OU process defined in the
Cartesian frame. Again all results are verified by computer
simulations.
%In Sec.~\ref{sec:overdamped} we finally consider the overdamped limit of
%the OU process, which reproduces to some extent the results derived in
%Sec.~\ref{sec-RW} for the simple random walk.
We conclude with a summary of our main results, a wider embedding and
an outlook to further research in Sec.~\ref{sec:concl}.

\section{\label{sec:level2}Stochastic Processes in Two Dimensions}

In this section we establish the basic analytical framework for
transforming two-dimensional stochastic processes from the Cartesian
into the comoving frame by introducing three distinct frames of
reference: the {\em Cartesian}, the {\em polar} and the {\em comoving}
frame. Figure~\ref{fig:eagle} shows a cartoon of an organism moving
along a time-discrete random trajectory in a plane. In the fixed
Cartesian frame, its velocity vector $\mathbf{v}$ is described by the
two components $(v_x, v_y)$, which specify the direction and magnitude
of motion relative to a stationary reference system. By contrast, the
comoving frame characterizes the same motion in terms of the speed $s$
and the turning angle $\alpha$, emphasizing changes in orientation
relative to the direction of motion.

\begin{figure}[hbt]
\centering
\includegraphics[width=1\columnwidth]{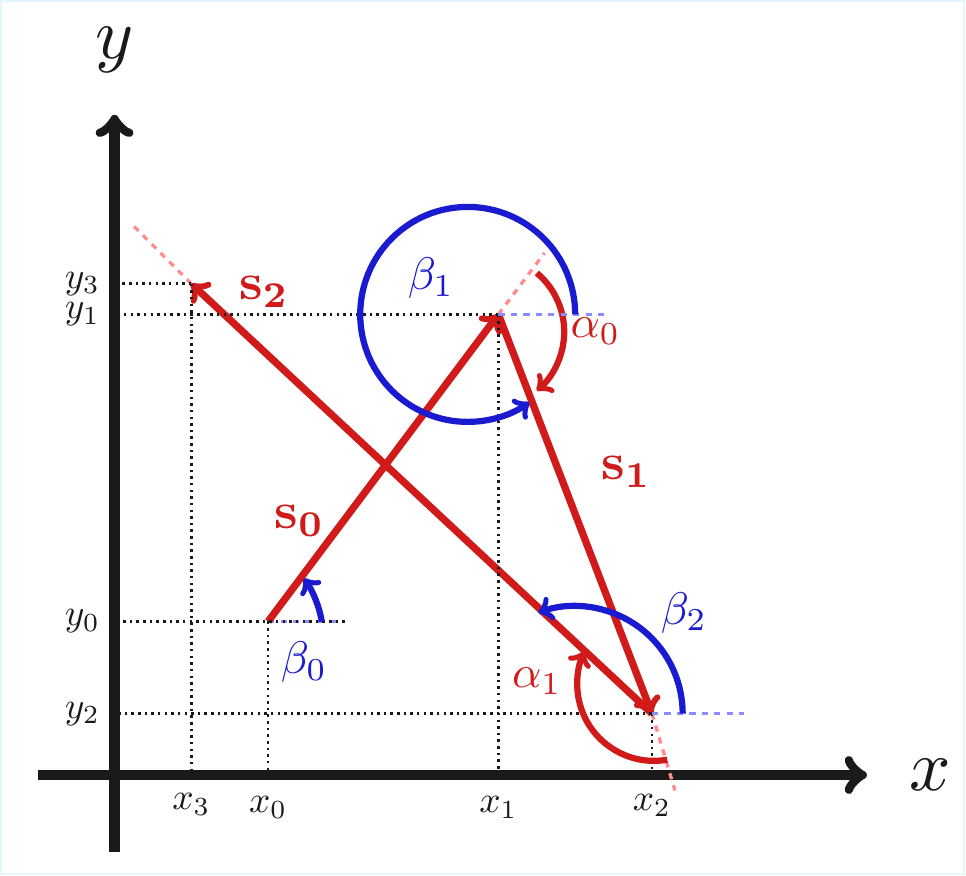}
\caption{Conceptual illustration of the interplay between three
  different reference frames for a process over four discrete time
  steps $n=0,1,2,3$: Cartesian frame described by positions $x_n$ and
  $y_n$ (black), polar frame described by the orientation angle $\beta_n$
  (blue) and the speed $s_n$ (red), and comoving frame described by the
  turning angle $\alpha_n$ and speed $s_n$ (red).}\label{lw2dm}
\end{figure}

Figure~\ref{lw2dm} depicts the geometric relationships between these
three reference frames by illustrating their different
representations. A clear understanding of the transformations between
Cartesian, polar, and comoving descriptions is essential for our
subsequent systematic analysis of stochastic dynamics in two
dimensions, and for interpreting the underlying physical and
biological processes across these different modeling contexts.
Extending the concept of stochastic processes evolving randomly over
time from one to two dimensions introduces greater complexity and
richness, due to considering a new degree of freedom. Mathematically,
within our context a time-discrete two-dimensional stochastic process
is typically represented by a vector $\mathbf{x}_n$ for the particle's
position coordinates at discrete time $n\in\mathbb{N}_0$ in the
Cartesian frame, characterized by probabilistic properties such as
distributions, mean, variance, and correlations in both components
$x_n$ and $y_n$. We now introduce these three reference frames in
detail and define the different transformations between them.

\subsection{Frames of reference and transformations}

\begin{enumerate}
\item \textbf{Cartesian Frame}: This fundamental reference frame uses
  position coordinates $(x_n, y_n)$ and velocity components $(v_{x,n},
  v_{y,n})$ for describing a particle's motion. Here and in the
  following we consider this frame to be at rest. For stochastic
  processes it is typically assumed that motion along the $x$-axis
  does not influence motion along the $y$-axis, and vice-versa,
  allowing for separate treatment of dynamics in each dimension. This
  simplifies the analysis for systems in terms of decoupled motion in
  different directions.

\item \textbf{Polar Frame}: This frame is based on a polar coordinate
  transformation of the particle's velocities, which is particularly useful for
  systems where radial and angular motion are key. Cartesian
  coordinates $(v_{x,n}, v_{y,n})$ are transformed into polar
  coordinates $(s_n, \beta_n)$, where $s_n$ is the absolute value of
  the vector velocity and $\beta_n$ is the polar angle of this
  vector. The transformation is trivially given by
\begin{eqnarray}
    s_n &= &\sqrt{v_{x,n}^2 +v_{y,n}^2}  \\
\beta_n &= &\arctan\left(\frac{v_{y,n}}{v_{x,n}}\right) \label{polcart-}.
\end{eqnarray}
Conversely, to convert back to Cartesian coordinates we have
\begin{align}
v_{x,n} &= s_n\cos \beta_n \\
v_{y,n} &= s_n\sin\beta_n .\label{cartpol}
\end{align}

\item \textbf{Comoving Frame}: This dynamic, particle-centered frame
  co-moves and -rotates with the particle, characterized by the
  particle's {\em speed} $s_n$ (as in Eq.(\ref{polcart-})) and {\em
    turning angle} $\alpha_n$. It is defined with respect to the
  change of the time-discrete orientation angle of the vector
  velocities as
\begin{align}
\alpha_n &= \beta_n-\beta_{n-1} \label{comcart}\: .
\end{align}
Here, $\alpha_n$ yields the turning angle between consecutive time
steps $n$ and $n-1$. The inverse transformations from the comoving
frame back to Cartesian velocity components are
\begin{align}
\beta_{n+1}&=\beta_n +\alpha_n \label{betaback}\\
v_{x,n} &= s_n \cos{\beta_n} \\
v_{y,n} &= s_n \sin{\beta_n} \label{comcartb}.
\end{align}
These equations reconstruct $v_{x,n}$ and $v_{y,n}$ from $s_n$ and
$\alpha_n$. The comoving frame is ideal for self-consistently
describing intrinsic stochastic fluctuations driving a self-propelled
particle, such as in active matter systems or organismic movement
models, as we will show in the following.
\end{enumerate}

\begin{figure*}[htbp]
\centering
\includegraphics[width=1.8\columnwidth]{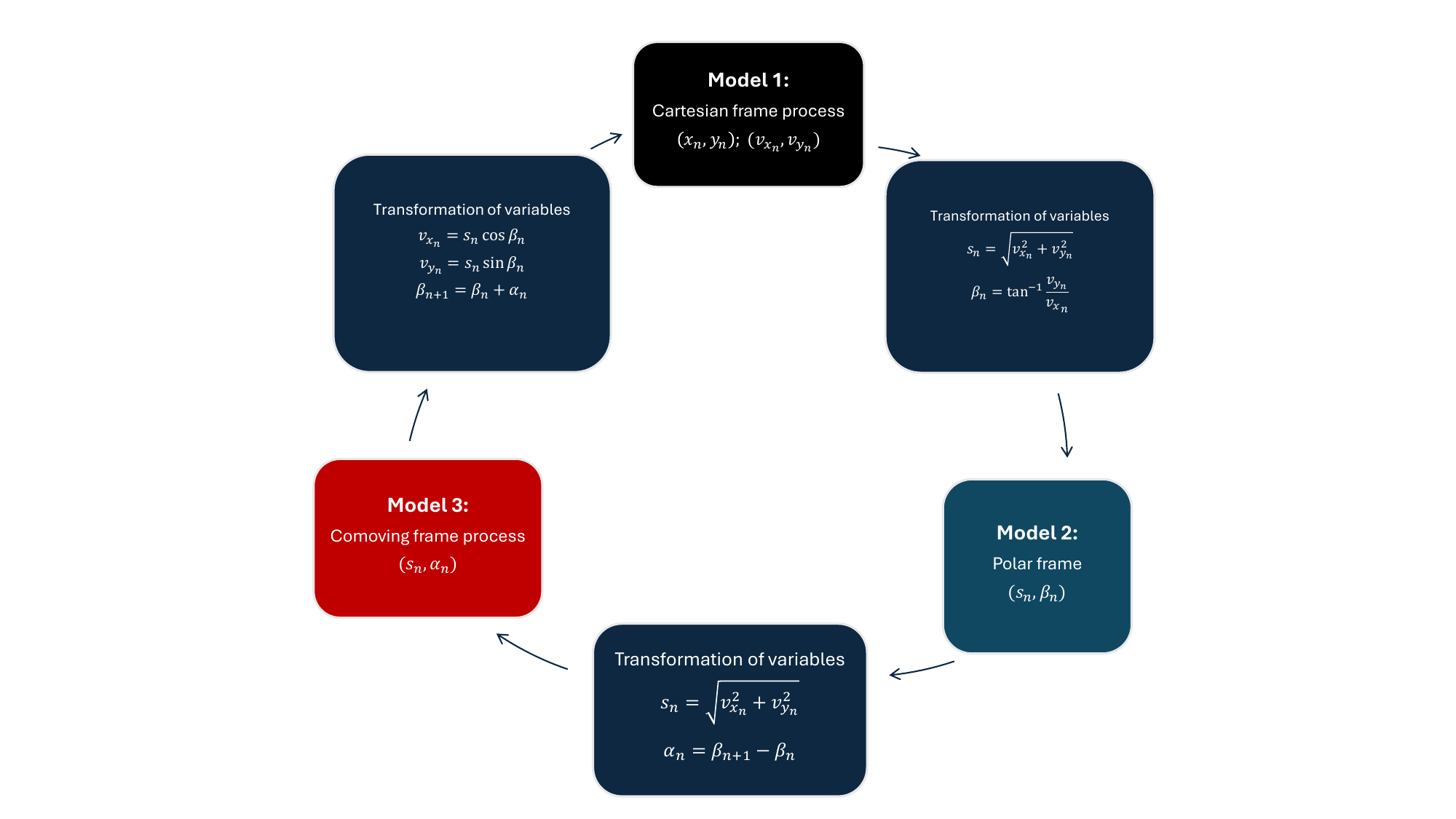}
\caption{Full transformation rules between the three different frames of reference and the respective inverse transfromation.}\label{trans-sum}
\end{figure*}

The interplay between these three different frames of reference in
terms of the respective transformations between them is summarised in
Fig.~\ref{trans-sum}.

\section{Three models for a simple Random Walk in two dimensions} \label{sec-RW}

In order to outline the basic principle of transforming a stochastic
process between these three different frames of reference, in the
following we first consider the example of a simple two-dimensional
random walk. We start by defining the random walk in Cartesian
coordinates. We then subsequently transform this process into polar
and then into comoving coordinates, hence arriving at three different
representations, or models, of this process in three different
frames. The analytical results are compared with computer
simulations. This analysis will pave the way to transform the OU
process, which we do in Sec.~\ref{sec-ou-comoving}.

\subsection{Model 1: Random Walk in the Cartesian frame} \label{sec:RW-cartesian}

We first consider a two-dimensional time-discrete random walk defined
by
\begin{equation}
    \mathbf{x}_{n+1} = \mathbf{x}_{n} + \Delta \mathbf{x}_n
\end{equation}
with increments $\Delta \mathbf{x}_n = \Delta t \mathbf{v}_n$, where
the discrete velocities $\mathbf{v}_n= (v_{x,n}, v_{y,n})$ are drawn
from a Gaussian distribution $P(v_x,\, v_y)$ with zero mean and
variance $\sigma^2$. This corresponds to an isotropic random walk in
the plane, with independent and identically distributed (i.i.d.)
steps.  The Cartesian formulation thus describes the random walk in
terms of Gaussian increments in each coordinate direction.

\subsection{Model 2: Random Walk in the polar frame} \label{sec:RW-polar}

For transforming to polar coordinates, we use the speed $s_n$ and
orientation angle $\beta_n$ defined by Eq.(\ref{polcart-}).  In order
to formulate a two-dimensional random walk in these coordinates, we
need to transform the Cartesian Gaussian velocity distribution into
polar coordinates. This change of variables is accomplished via using
conservation of probability and the Jacobian determinant yielding
\begin{equation}
    P(s,\beta) = s \, P(v_x,\, v_y),
\end{equation}
which in the isotropic Gaussian case of a normal distribution,
$v_x,v_y \sim \mathcal{N}(0,\sigma^2)$, becomes
\begin{equation}
    P(s,\beta) = \frac{s}{2\pi\sigma^2} 
    \exp\!\left(-\frac{s^2}{2\sigma^2}\right) \: , \: s \ge 0 \: , \: \beta \in (0,2\pi] .
\end{equation}
This factorizes into a Rayleigh distribution for the speed,
\begin{equation}
    P(s) = \frac{s}{\sigma^2}\exp\!\left(-\frac{s^2}{2\sigma^2}\right)
    \label{rw-s-pdf}
\end{equation}
and a uniform distribution for the polar angle,
\begin{equation}
    P(\beta) = \frac{1}{2\pi}.\label{rw-b-pdf}
\end{equation}
The autocorrelations of the polar variables vanish at nonzero lag, as
shown in more detail in Appendix \ref{app:no_autocorr},
\begin{align}
    \langle s_i s_j \rangle - \langle s_i \rangle \langle s_j \rangle &= 0, \\
    \langle \beta_i \beta_j \rangle &= 0,  \qquad i \neq j,
\end{align}
Thus, the polar representation naturally encodes the random walk in
terms of i.i.d.\ random variables $(s_n,\beta_n)$ sampled from a
Rayleigh-distributed speed and a uniformly distributed orientation
angle.

\subsection{Model 3: Random Walk in the comoving frame} \label{sec:RW-comoving}

The random walk can also be described in the comoving frame, where the
relevant variables are the speed $s_n$ and the turning angle
$\alpha_n$.  Since the orientation angles $\beta_n$ are uniformly
i.i.d.\ variables, the increments $\alpha_n$ defined by the linear
Eq.(\ref{comcart}) are also i.i.d.\ and uniformly distributed on the
circle, see Appendix \ref{app:no_autocorr} for more details. These
properties support a natural representation of the random walk in the
comoving frame, defined by the pair \((s_n,\alpha_n)\) by sampling the
speed i.i.d.\ as before from a Rayleigh distribution,
\begin{equation}
    s \sim P(s) = \frac{s}{\sigma^2}\exp\!\left(-\frac{s^2}{2\sigma^2}\right), \label{rw-s-pdf}
\end{equation}
and the turning angle \(\alpha\) i.i.d.\ from a uniform distribution
on the circle, as originally suggested in Ref.~\cite{Pea05},
\begin{equation}
    \alpha \sim P(\alpha) = \frac{1}{2\pi}, \label{rw-ta-pdf}
    \qquad \alpha \in (-\pi,\pi].
\end{equation}
Here and in the following we map $\alpha$ onto the circle by applying
a modulo $2\pi$ operation and restricting it to the interval
$(-\pi,\pi]$. This choice corresponds to the minimal angular
  displacement and captures the tendency of an organism to change
  direction through the shortest possible rotation \cite{Gomez-Solano_2020}. It
  also avoids a subtle issue of chirality in case we only admitted
  positive turning angles. The issue of wrapping onto the circle
  becomes more subtle for general stochastic processes, as we will
  discuss for the example of the OU process in
  Sec.~\ref{sec-ou-comoving}. Thus, in the comoving frame, the random
  walk is governed again by two independent noise sources: Rayleigh
  noise in the speed and uniform white noise in the turning angle.  A
  compact summary of the probability distributions and autocorrelation
  properties of all relevant variables in the three models (Cartesian,
  polar and comoving) is provided in Table~\ref{tab:rw-properties}.

\begin{figure}[ht!]
\centering
  \includegraphics[width=6cm, height=5cm]{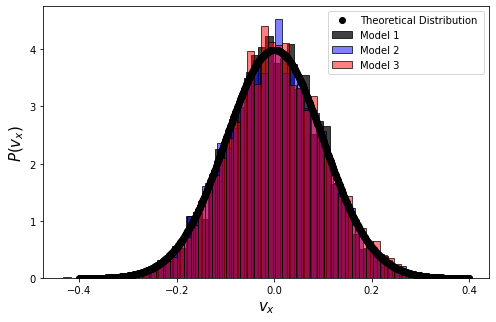}  
  \includegraphics[width=6cm, height=5cm]{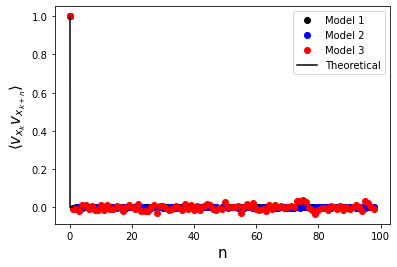} 
  \caption{
    Top: Simulation results for the probability density of the
    velocity component $v_x$ for the simple random walk formulated in
    three different frames, yielding three different random walk
    models, compared with the Gaussian distribution. Bottom: The
    autocorrelation function of the random walk velocities are all
    uncorrelated .}
  \label{velxpdfrw}
\end{figure}

\begin{figure}[ht!]
\centering 
  \includegraphics[width=6cm, height=5cm]{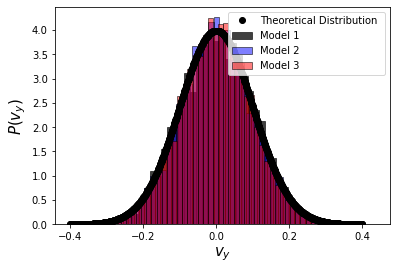} 
  \includegraphics[width=6cm, height=5cm]{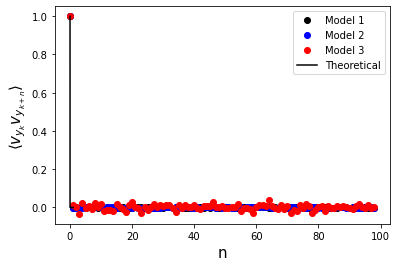} 
  \caption{Same as Fig.~\ref{velxpdfrw} for the velocity component
    $v_y$}
  \label{velypdfrw}
\end{figure}

\begin{figure}[ht!]
\centering
  \includegraphics[width=6cm, height=5cm]{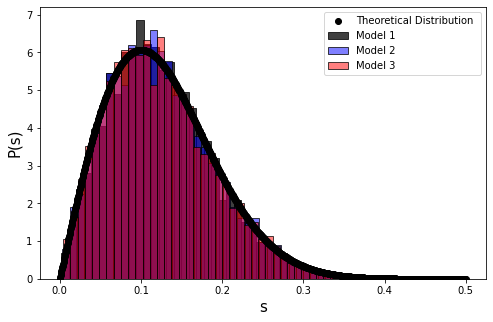} 
  \includegraphics[width=6cm, height=5cm]{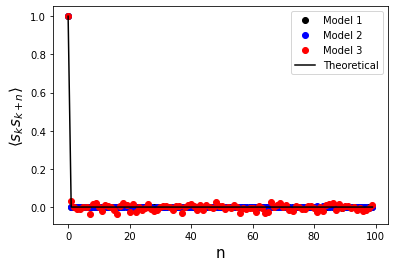} 
   
  \caption{Top: Simulation results for the probability density of the
    speed $s$ for the simple random walk in the three different frames
    compared with Eq.(\ref{rw-s-pdf}). Bottom: The autocorrelation
    function of the speed is again uncorrelated.}
  \label{speedrw}
\end{figure}

\begin{figure}[ht!]
\centering
  \includegraphics[width=6cm, height=5cm]{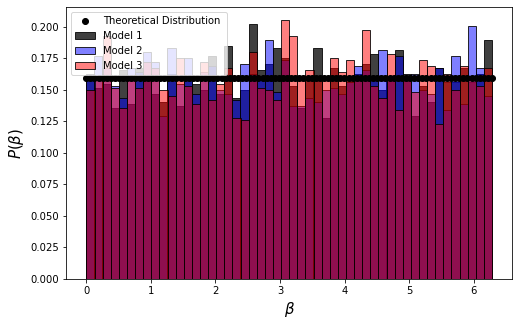} 
  \includegraphics[width=6cm, height=5cm]{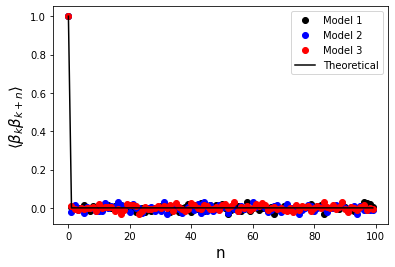} 
  
  \caption{Top: Probability density of the orientation angle $\beta_n$
    for the simple random walk in the three different frames compared
    with Eq.(\ref{rw-b-pdf}). Below: The autocorrelation function is
    again uncorrelated.}
  \label{betarw}
\end{figure}

\begin{figure}[ht!]
\centering
  \includegraphics[width=6cm, height=5cm]{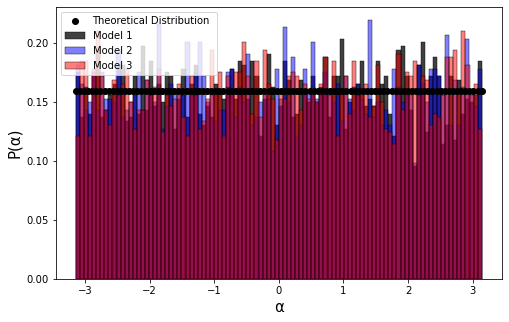} 
  \includegraphics[width=6cm, height=5cm]{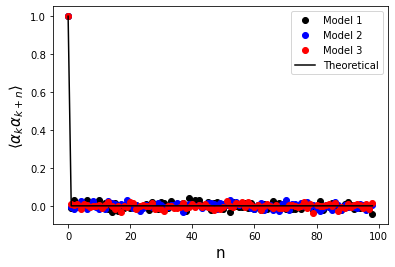} 
  
  \caption{Top: Probability density of the turning angle $\alpha_n$
    for the simple random walk in the three different frames compared
    with Eq.(\ref{rw-ta-pdf}). Bottom: The autocorrelation function is
    again uncorrelated.} 
  \label{tarw}
\end{figure}

In summary, for a stochastic process defined in the Cartesian frame
one needs to transform the corresponding probability distributions
between the different frames, and in general to check for the
correlation functions of the associated variables in the different
frames (even though here for the Markovian random walk there is little
to check). These analytical calculations are supported by numerical
simulations.  Figures~\ref{velxpdfrw},\ref{velypdfrw} show that the
probability densities of the Cartesian velocity components $v_x$ and
$v_y$ are perfectly Gaussian for all three stochastic models, while
the lower panels confirm that their autocorrelations vanish beyond
zero lag, consistent with $\delta$-correlated increments.
Figure~\ref{speedrw} compares the simulated probability densities of
the speed with the analytical Rayleigh distribution
Eq.~\eqref{rw-s-pdf} and demonstrates the absence of temporal
correlations for all three models.  Similarly, Fig.\ref{betarw}
shows the uniform distribution of the orientation angle $\beta$
Eq.~\eqref{rw-b-pdf} and its lack of autocorrelation for all three
models. Finally, Fig.~\ref{tarw} illustrates the distribution of
turning angles $\alpha$, which follows the uniform law in
Eq.~\eqref{rw-ta-pdf}, and confirms its interpretation as a white
noise process in angular space for all three models.

\begin{table*}[ht]
\centering
\caption{Statistical properties of different two-dimensional random
  walk variables in three different coordinate frames: Cartesian
  (Model~1), Polar (Model~2), and Comoving (Model~3).}
\label{tab:rw-properties}
\begin{ruledtabular}
\begin{tabular}{c>{\raggedright\arraybackslash}p{5.5cm}c}

Model \& Variables & Probability distributions & Autocorrelations  \\ \hline \\

\textbf{Model 1: Cartesian random walk} & & \\ 
Velocity components $(v_x,v_y)$ 
&$P(v_x,v_y) = 
\frac{1}{2\pi\sigma^2}\exp\!\left(-\frac{v_x^2+v_y^2}{2\sigma^2}\right)$ 
& $\langle v_{x,i}v_{x,j}\rangle = \langle v_{y,i}v_{y,j}\rangle = 0$ for $i\neq j$ \\[1em]

\textbf{Model 2: Polar random walk} & & \\
Speed $s=\sqrt{v_x^2+v_y^2}$ 
& $P(s) = \frac{s}{\sigma^2}\exp\!\left(-\frac{s^2}{2\sigma^2}\right), \quad s\ge 0$ 
& $\langle s_i s_j \rangle - \langle s_i \rangle \langle s_j \rangle = 0$ for $i\neq j$ \\

Orientation angle $\beta = \arctan(v_y,v_x)$ 
& $P(\beta) = \frac{1}{2\pi}, \quad \beta \in (0,2\pi]$ 
& $\langle \beta_i \beta_j \rangle = 0$ for $i\neq j$ \\[1em]

\textbf{Model 3: Comoving random walk} & & \\
Speed $s$ 
& $P(s) = \frac{s}{\sigma^2}\exp\!\left(-\frac{s^2}{2\sigma^2}\right), \quad s\ge 0$ 
& $\langle s_i s_j \rangle - \langle s_i \rangle \langle s_j \rangle = 0$ for $i\neq j$\\

Turning angle $\alpha = \beta_i - \beta_{i-1}$ 
& $P(\alpha) = \frac{1}{2\pi}, \quad \alpha \in (-\pi,\pi]$ 
& $\langle \alpha_i \alpha_j \rangle = 0$ for $i\neq j$ \\[1em]

\end{tabular}
\end{ruledtabular}
\end{table*}



\section{Three Models for Ornstein-Uhlenbeck in two dimensions}\label{sec-ou-comoving}

The conceptual framework of how to transform between the three
different frames, illustrated in the previous section for the simple
Markovian random walk, is now rolled out for the OU process yielding
the velocities of random movements in a plane.
%\subsection{The Ornstein-Uhlenbeck Process in One Dimension}\label{sec:1dOU}
Before elaborating on OU dynamics \cite{OU1930} in two dimensions, let
us briefly recall its one-dimensional formulation and key statistical
properties. The OU process is defined by the underdamped Langevin
equation
\begin{equation}
    dv(t) = -\frac{\gamma}{m} v(t)\, dt + \sqrt{2\tilde{D}}\, dW_t, \label{eq:1dOU}
\end{equation}
where $dW_t$ is the increment of a Wiener process, $\gamma/m$ is the
relaxation rate, and the diffusion coefficient $\tilde{D}$ is related
to the conventional diffusion coefficient $D$ for the position via $D
\to \frac{\tilde{D} m^2}{\gamma^2}$ \cite{dybiec2024multimodality}.
The solution of Eq.~\eqref{eq:1dOU} is a Gaussian process with mean
\begin{equation}
    \langle v(t) \rangle = v(0) e^{-\frac{\gamma}{m}t} \label{eq:1dOUmean}
\end{equation}
and variance
\begin{equation}
    \text{Var}[v(t)] = \frac{\tilde{D} m}{\gamma}\left(1-e^{-2\frac{\gamma}{m}t}\right). \label{eq:1dOUvar}
\end{equation}
In the long-time limit, the process reaches a stationary state with Gaussian distribution
\begin{equation}
    P_{\text{st}}(v) = \sqrt{\frac{\gamma}{2\pi \tilde{D} m}}
    \exp\!\left(-\frac{\gamma v^2}{2\tilde{D}m}\right), \label{eq:1dOUstat}
\end{equation}
which has zero mean and variance $Dm/\gamma$.
The velocity autocorrelation function of the stationary process is
\begin{equation}
    \langle v(t)v(t') \rangle = \frac{\tilde{D} m}{\gamma} \, e^{-\frac{\gamma}{m}|t-t'|}, \label{1doucorr}
\end{equation}
showing the characteristic exponential decay with correlation time
$\tau=m/\gamma$. The one dimensional OU process is therefore {\em
  Gaussian} (all finite-dimensional distributions are Gaussian), {\em
  stationary} (the distribution converges to the steady Gaussian
Eq.\eqref{eq:1dOUstat}) and {\em Markovian} (the process has the Markov
property, enabling Itô transformations to other coordinate systems
\cite{Gard09}). These three properties make OU dynamics the
prototypical model of a stochastic process with exponential
memory. The two-dimensional OU process in Cartesian coordinates is
then constructed as two independent copies of Eq.~\eqref{eq:1dOU}, one
for each velocity component.

\subsection{Model 1: Cartesian Two-Dimensional OU Process}
\label{sec:model1}

We first consider a two-dimensional OU-driven walk in the Cartesian
frame, constructed from two independent one-dimensional OU processes
for the velocities along the $x$ and $y$ axes. The velocity components
$\mathbf{v}=(v_x,v_y)$ thus evolve according to the stochastic
differential equation
\begin{equation}
d\mathbf{v}=-\frac{\gamma}{m}\mathbf{v}\, dt + \sqrt{2\tilde{D}}\, \xi_{\mathbf{v}}(t)\, dt, \label{sde2ouv}
\end{equation}
where $\xi_{\mathbf{v}}=(\xi_{v_x},\xi_{v_y})$ are independent Gaussian white noise terms.   

The OU process exhibits exponentially correlated velocities with correlation function
\begin{equation}
\langle v_j(t)v_j(t')\rangle - \langle v_{0_j}\rangle^2 = \frac{\tilde{D} m}{\gamma}  e^{-\frac{\gamma}{m}|t-t'|} \label{ACV}
\end{equation}
with $j \in (x,y)$.  The stochastic differential equation
\eqref{sde2ouv} is solved numerically in equilibrium using the
Euler–Maruyama method \cite{Kloeden92} with parameters $D=\gamma=m=1$
and noise terms $\xi_{\mathbf{v}}$ of zero mean and unit standard
deviation. After integration, the corresponding polar frame variables
are obtained via the polar transformation
\begin{equation}
s = \sqrt{v_x^2 + v_y^2}, \quad \beta = \arctan(v_y,v_x), \label{polcart}
\end{equation}
where $s$ is the instantaneous speed and $\beta$ the orientation angle.  

We recall that the orientation angle $\beta$, extracted via the
arctangent function, is defined modulo $2\pi$, with values taken in
the interval $[0,2\pi)$. Consequently, the turning angle $\alpha
  =\Delta \beta$, as a difference between these orientations,
  initially takes values on the full real line
  $(-\infty,\infty)$. Interestingly, this unwrapped turning angle
  (prior to any modulo operation) exhibits non-trivial temporal correlations. Wrapping $\alpha$ onto the
  circle, as by definition needed for comoving frame
  dynamics, eliminates these correlations, thus effectively
  Markovianising the turning angle dynamics. This represents a loss of
  information in the comoving frame about the history of previous
  rotations with respect to the Cartesian frame.

  %likely does not matter because of circular symmetry of the process...
  %
  %Through this construction, all variables are consistently
  %defined across the three reference frames considered.

\subsection{Model 2: Polar OU Process Driven by Itô Equations}
\label{sec:model2}

Defining stochastic equations in the polar frame introduces an
Itô–Stratonovich ambiguity, as the transformation from Cartesian
velocities to polar coordinates generates multiplicative noise for the
speed $s$ and orientation $\beta$.
%The objective of this model is to
%derive and simulate such equations to reproduce the dynamics of the
%two-dimensional Cartesian OU process entirely within the comoving
%frame.
Using Itô calculus \cite{Gard09}, Eq.~\eqref{sde2ouv} can be
transformed into two coupled stochastic differential equations for the
speed $s$ and orientation $\beta$ (see Appendix \ref{AITO}), as
similarly derived in Ref.~\cite{Gard09} for an electric field, and for
the Stratonovich approach in Ref.~\cite{RBELS12}:
\begin{align}
\frac{d\beta}{dt} &= \frac{1}{s}\sqrt{2\tilde{D}}\,\xi_{\beta}(t) \label{eq:itop} \\
\frac{ds}{dt} &= \left(-\frac{\gamma}{m}s + \frac{\tilde{D}}{s}\right) + \sqrt{2\tilde{D}}\,\xi_{s}(t), \label{eq:itos}
\end{align}
where the noise terms are defined as
\begin{align}
\xi_{\beta}(t) &= -\xi_x(t)\sin\beta + \xi_y(t)\cos\beta \\
\xi_{s}(t) &= \xi_x(t)\cos\beta + \xi_y(t)\sin\beta
\end{align}
with $\xi_x$ and $\xi_y$ being the independent Gaussian white noises
in the Cartesian frame. It can be shown that $\xi_\beta$ and $\xi_s$
are also Gaussian white noises \cite{Gard09}.  The corresponding
Fokker–Planck equation \cite{Risk} for the joint probability density function
$P(s,\beta,t)$ in the Itô interpretation is 
\begin{align}
\frac{\partial P(s,\beta,t)}{\partial t} &= -\frac{\partial}{\partial s} \left( -\frac{\gamma}{m}s + \frac{\tilde{D}}{s} \right) P(s,\beta,t) \nonumber \\
&\quad + \tilde{D} \frac{\partial^2 P(s,\beta,t)}{\partial s^2} + \frac{\tilde{D}}{s^2} \frac{\partial^2 P(s,\beta,t)}{\partial \beta^2}, \label{FPITO}
\end{align}
which admits the stationary solution
\begin{equation}
P(s,\beta) = \mathcal{N}\, s\, e^{-\frac{\gamma s^2}{2 \tilde{D} m}}, \label{jpdou}
\end{equation}
where $\mathcal{N}$ represents the uniform distribution of $\beta$ and
the speed $s$ follows a Rayleigh distribution.
%Numerical simulations of Eqs.~\eqref{eq:itop} and \eqref{eq:itos} confirm that this polar OU formulation accurately reproduces the statistics of the Cartesian OU process.
For simulations, Eqs.~\eqref{eq:itop} and \eqref{eq:itos} are
integrated using the Euler–Maruyama method \cite{Kloeden92}, ensuring
that the standard deviations of $\xi_s$ and $\xi_\beta$ match those of
the Cartesian noise in Model 1. A key consideration arises from
Eq.~\eqref{eq:itos}: Numerically, the Gaussian noise in $\xi_s$ can
occasionally produce negative values for the speed. To maintain
positive speeds, any negative speed values encounter in the simulation, are replaced by a small
threshold value of $10^{-2}$. Note that analytically negative speeds Eq.~\eqref{eq:itos}
appear to be eliminated by the Itô $1/s$ flux term, although we have
no proof of the positivity of the speed in this equation.

The autocorrelation function of the speed can be calculated (see
Appendix \ref{acs}) as
\begin{equation}
\langle s(t)s(t')\rangle - \langle s(t)\rangle^2 = \frac{2 \tilde{D} m}{4 \gamma} \, e^{-2\frac{\gamma}{m}|t-t'|}, \label{acfs}
\end{equation}
which is analogous to the velocity correlations in the Cartesian frame
Eq.~\eqref{ACV} but decays at twice the rate, $2\gamma/m$, reflecting
the distinct dynamics in polar coordinates.

Deriving a closed-form expression for the autocorrelation of $\beta$
is more involved due to its inverse dependence on $s$, which
introduces a nontrivial coupling between speed and
orientation. Consequently, instead we consider the autocorrelation of
$\cos \beta$, which can be approximated as discussed in
Appendix~\ref{AP-beta} to obtain 
\begin{align}
\langle \cos\beta(t)\cos\beta(t+\tau)\rangle
\approx
\frac{1}{2}\,
\exp\!\Big[
-\tilde D\,\tau\,
\Big\langle s^{-2}\Big\rangle
\Big].\label{approx_beta_result}
\end{align}

\subsection{Model 3: OU Process in the Comoving Frame}
\label{sec:model3}

\begin{figure}[ht!]
\centering
\includegraphics[width=6cm, height=4cm]{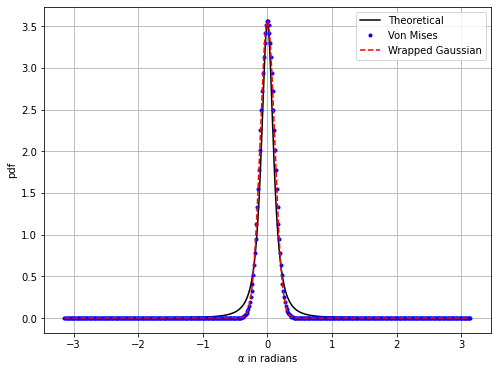}
\caption{Theoretical probability distribution for the turning angle
  $\alpha$, Eq.~\eqref{turpdf}, compared with a wrapped Gaussian and a
  von Mises distribution. One can see clear deviations in the tails.}
\label{von-wrap-the}
\end{figure}

To achieve a self-consistent description of the two-dimensional OU
process in the comoving frame, it is necessary to formulate an
equation for the turning angle $\alpha$. This can be accomplished by
using Eq.~\eqref{comcart} and computing the corresponding distribution
of $\alpha$ from the right-hand side of Eq.~\eqref{eq:itop},
restricting the turning angle to the interval $[-\pi, \pi]$ with the
modulo $2\pi$ operation. The resulting turning angle probability
distribution is given accordingly to (see Appendix \ref{ap:tapdf})
\begin{equation}
P(\alpha) = \frac{\sqrt{2\tilde{D}\Delta t}}{2\pi} \sum_{k=-\infty}^{\infty} \left|\frac{\sigma}{\sigma_R} k\right| K_1\left( \left|\frac{\sqrt{2\tilde{D}\Delta t}\sigma k}{\sigma_R}\right|\right) e^{i k \alpha}, \label{turpdf}
\end{equation}
where $\Delta t$ is the discretisation time for the originally
time-continuous Eq.~\eqref{comcart}, and $\sigma,\sigma_R$ are the
standard deviations of the Gaussian white noise $\xi_\beta$ (from
Eq.~\eqref{eq:itop}) and the exponentially correlated speed $s(t)$
(from Eq.~\eqref{jpdou}), respectively. Here, $K_1(z)$ denotes the
modified Bessel function of the second kind. While $P(\alpha)$
resembles a wrapped Gaussian or von Mises distribution as shown in
Fig.~\ref{von-wrap-the}, its precise functional form defines, to our
knowledge, a new distribution for the turning angle.

In a steady state, the speed $s$ is distributed according to a
Rayleigh distribution with exponentially decaying autocorrelations as
in Eqs.~\eqref{jpdou},\eqref{acfs}. This leads to a self-consistent
formulation of a Cartesian OU process represented in the comoving
frame as
\begin{equation}
\alpha_n = \xi_\alpha(n\Delta t), \quad s_n = \tilde{\xi}_s(n\Delta t), \label{eq:oucm}
\end{equation}
where $\xi_\alpha$ is white noise with the functional form given by
$P(\alpha)$, and $\tilde{\xi}_s$ is exponentially correlated Rayleigh
noise. Alternatively, the speed can be generated using
Eq.~\eqref{eq:itos} with Gaussian white noise. What we mean with a
self-consistent description of the OU process in the comoving frame is
that the above two equations are formulated solely in terms of speed
and turning angle, without explicitly involving any other
coordinates. This is not the case in the polar representation, which
still somewhat relies on the Cartesian frame. Note that, exactly in
this sense, here we go beyond previous formulations of active particle
models, such as in Refs.~\cite{RoSch11,GSR12,RBELS12,BeDiL16}, where
the active part is expressed in terms of polar coordinates but not defined
self-consistently by using the turning angle as a comoving frame
coordinate, as above. In principle, for a walker, or an organism,
knowledge of a sequence of speed and turning angle is fully sufficient
to self-generate movement, which is the idea underlying Pearson's
original random walk and the CRW of ME. However, to our knowledge,
generalised processes of the form of Eq.~\eqref{eq:oucm} have not yet been
considered for modelling organismic movements.

For simulations, the variables in the comoving frame can be generated
directly from their respective distributions, i.e., the Markovian
turning angle $\alpha$ is sampled i.i.d.\ from $P(\alpha)$
Eq.(\ref{turpdf}). However, to transform back to the polar or
Cartesian frames by using Eq.(\ref{betaback}) for the orientation
angle $\beta$, the correct time discretisation taking the diffusive
stochastic scaling into account has to be considered \cite{Gard09}. For this we introduce an additional helper
variable, the angular velocity $\omega = d\beta/dt$, thus writing
$\alpha_n \sim \omega\sqrt{\Delta t}$, where $\omega$ is sampled from
the probability distribution for the angular velocity associated with
the turning angle distribution, see Eq.~\eqref{pdf-angularv} in
Appendix~\ref{ap:tapdf}. Simultaneously, the speed $s$ is sampled from exponentially correlated
Rayleigh noise. This procedure maintains the correct temporal
correlations and steady-state statistics of the Cartesian OU
process. Correlations for $s$ can be introduced numerically using the
Cholesky decomposition method \cite{choleskynumpy}, generating two
Gaussian variables with exponential correlations as in
Eq.~\eqref{ACV}. According to \cite{Hogema2005}, if $X$ and $Y$ are
independent Gaussian variables, then $R = \sqrt{X^2 + Y^2}$ follows a
Rayleigh distribution with standard deviation $\sigma_R$. In this
manner, exponentially correlated Rayleigh noise is obtained. Along
these lines, once the comoving variables $(\alpha, s)$ are generated,
the transformations in Eq.~\eqref{comcart}, modulo using $\omega$, are
applied to recover the Cartesian velocities.

\begin{figure}[ht!]
\centering
\includegraphics[width=6cm, height=4cm]{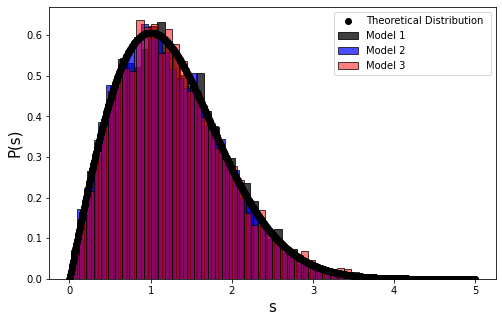}
\includegraphics[width=6cm, height=4cm]{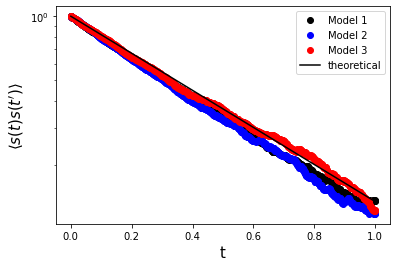}
\caption{Top: Probability distributions for the speed $s$ in the three
  different models of an OU process in two dimensions, compared with
  Eq.(\ref{jpdou}) (lines). Bottom: Corresponding speed
  autocorrelations compared with Eq.(\ref{acfs}) for the three
  different models. Here and in the following $t$ represents the total time length of the process.} \label{sim.3mod-s}
\end{figure}

\begin{figure}[ht!]
\centering
\includegraphics[width=6cm, height=4cm]{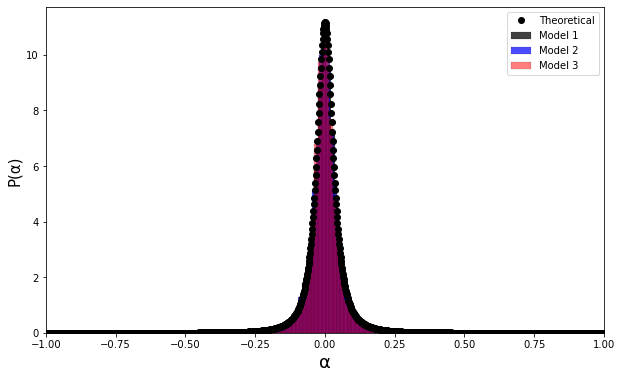}
\includegraphics[width=6cm, height=4cm]{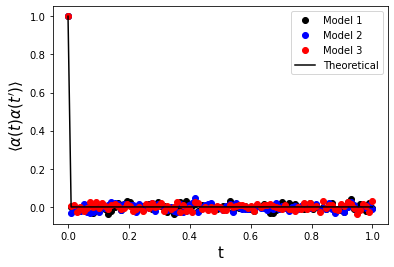}

\caption{Top: Probability distributions for the turning angle $\alpha$
  in the three different models of an OU process in two dimensions,
  compared with Eq.(\ref{turpdf}) (lines), the original distribution is defined in the interval $[-\pi,\pi)$, but a focus is made in the interval $[-1,1]$ for visual purposes. Bottom: Corresponding
  autocorrelations for the three different models, compared with a
  $\delta$ function.} \label{sim.3mod-a}
\end{figure}

\begin{figure}[ht!]
\centering
\includegraphics[width=6cm, height=4cm]{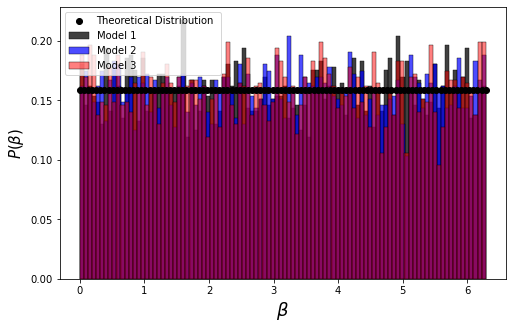}
\includegraphics[width=6cm, height=4cm]{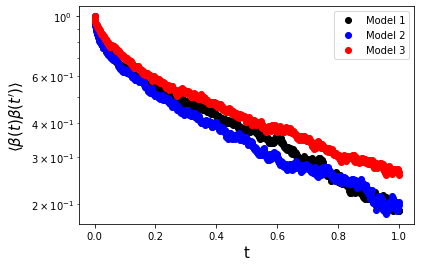}
\includegraphics[width=6cm, height=4cm]{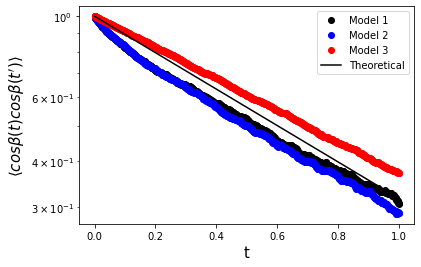}
\caption{Top: Probability distributions for the orientation angle
  $\beta$ in the three models, compared to a uniform
  distribution. Middle: Corresponding orientation autocorrelations for
  the three models. Bottom: Corresponding autocorrelations for the
  cosine of the orientation, including the analytical approximation
  Eq.~\eqref{approx_beta_result} (line).} \label{sim.3mod-b}
\end{figure}

\begin{figure}[ht!]
\centering
\includegraphics[width=6cm, height=4cm]{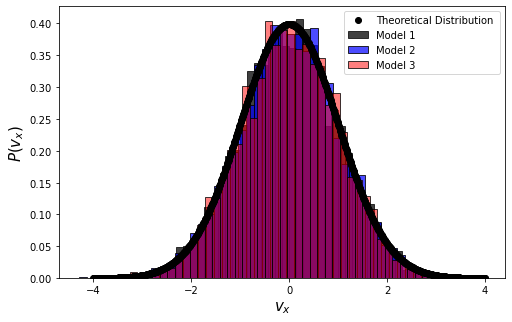}
\includegraphics[width=6cm, height=4cm]{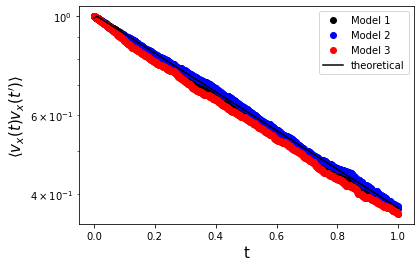}
\caption{Top: Probability distributions for the velocity $v_x$ in the
  three models compared with the corresponding theoretical expression
  for a Gaussian distribution (line). Bottom: Correspondind velocity
  autocorrelations compared with the theoretical expression (line)
  in Eq.(\ref{1doucorr}).} \label{sim.3mod-v}
\end{figure}

This description completes our definition of the three models of the
OU process in the three different coordinate frames. We now compare
results from simulating these three models with each other, and with
analytical results, to demonstrate that all these different
formulations reproduce the same statistical properties in the different
frames. The simulation results presented in Figs.~\ref{sim.3mod-s},
\ref{sim.3mod-a}, \ref{sim.3mod-b}, and
\ref{sim.3mod-v} show the probability distributions and associated
autocorrelation functions for each key variable: speed $s$, turning
angle $\alpha$, orientation angle $\beta$, angular velocity $\omega$,
and velocities $v_x$, $v_y$, as summarized in Table
\ref{tab:ou-properties-sdes}. The figures demonstrate strong agreement
between all three models. The probability distributions for each
variable exhibit very similar shapes and peak locations, indicating
that the underlying statistical properties of the simulated
trajectories are well preserved. Likewise, the autocorrelation
functions decay at comparable rates, demonstrating that temporal
correlations and memory effects are consistently captured across the
different model implementations. For instance, Fig.\ref{sim.3mod-s}
presents the speed distributions and autocorrelations in the comoving
frame, which closely match the theoretical Rayleigh distribution and
the exponential decay predicted by
Eq.(\ref{acfs}). Figure \ref{sim.3mod-a} shows the turning angle
distributions compared with Eq. (\ref{turpdf}) and their approximately
$\delta$-like autocorrelations.
%, while Fig.(\ref{sim.3mod-av}) illustrates the behavior of the angular velocity $\omega$. 
The orientation angle $\beta$ distributions and correlations are displayed
in Fig.\ref{sim.3mod-b}. We remark again that obtaining an analytical
expression for the autocorrelation of $\beta$ looks impossible due to
its dependence on the speed in Eq.(\ref{eq:itop}). Instead, we
therefore consider our analytical approximation for the
autocorrelation of $\cos \beta$. A small mismatch is observed in these
results, which can be attributed to two main factors: (i) when
sampling the random variables in the comoving frame from
Eq. (\ref{turpdf}), the infinite discrete Fourier series must be
truncated at a finite value $N$ due to the computational cost of
simulating large terms. This truncation directly affects the accuracy
of the sampled variables. (ii) The choice of the simulation step size
$\Delta t$ influences the recovery of continuous trajectories, i.e.,
smaller steps improve the accuracy but increase computational expense.
Finally, the Cartesian velocity components $v_x$ and $v_y$,
illustrated in Fig.\ref{sim.3mod-v}, confirm Gaussian behavior with
temporal correlations consistent with theoretical predictions, further
validating the equivalence of the constructed three models.
%\begin{figure}[ht!]
%\centering
%\includegraphics[width=6cm, height=4cm]{Figs/av pdf.png}
%\includegraphics[width=6cm, height=4cm]{Figs/av corr.png}
%\caption{Top: Probability distributions for the angular velocity
 % $\omega$ in the three models, compared with
  %Eq.(\ref{pdf-angularv}) (line). Bottom: Corresponding
  %autocorrelations for the 3 different models, compared with a delta
  %function.} \label{sim.3mod-av}
%\end{figure}

\begin{table*}[ht]
\centering
\caption{Statistical properties of different two-dimensional OU
  process variables in three different coordinate frames: Cartesian
  (Model~1), Polar (Model~2), and Comoving (Model~3).}
\label{tab:ou-properties-sdes}
\renewcommand{\arraystretch}{1.4}
\begin{ruledtabular}
\begin{tabular}{ccc}

  Model \& Variables & Probability distributions & Autocorrelations \\ \hline \\
  
  \textbf{Model 1: Cartesian OU} & & \\
Velocity components $(v_x,v_y)$ 
\\$d\mathbf{v}=-\frac{\gamma}{m}\mathbf{v} dt + \sqrt{2\tilde{D}}\,\xi_{\mathbf{v}}(t) dt$ 
& $P(v_i) = \frac{1}{\sqrt{2\pi \sigma_i^2}}\exp\left(-\frac{v_i^2}{2\sigma_i^2}\right)$ 
& $\langle v_j(t)v_j(t')\rangle - \langle v_{0_j}\rangle^2 = \frac{\tilde{D} m}{\gamma}  e^{-\frac{\gamma}{m}|t-t'|}$\\[0.8em]

\textbf{Model 2: Polar OU} & & \\
Speed $s$ 
\\  $\dot{s}=-\frac{\gamma}{m}s + \frac{\tilde{D}}{s} + \sqrt{2\tilde{D}}\,\xi_s(t)$
& $P(s) = s\, e^{-\frac{\gamma s^2}{2\tilde{D} m}}, \quad s\ge 0$ 
& $\langle s(t)s(t')\rangle - \langle s\rangle^2 = \frac{2\tilde{D} m}{4\gamma} e^{-2\frac{\gamma}{m}|t-t'|}$ \\[0.5em]

Orientation angle $\beta$ 
\\ $\dot{\beta} = \frac{1}{s} \sqrt{2\tilde{D}}\,\xi_\beta(t)$
& $P(\beta) = \frac{1}{2\pi}, \quad \beta \in [0,2\pi]$ 
& $\langle \cos\beta(t)\cos\beta(t+\tau)\rangle
\approx
\frac{1}{2}\,
\exp\!\Big[
-\tilde D\,\tau\,
\Big\langle s^{-2}\Big\rangle
\Big]$ \\[0.5em]

%Angular velocity $\omega$ 
%& $P(\omega) = \frac{\sqrt{2\tilde{D}}}{2\pi} \sum_{k=-\infty}^{\infty} |k| K_1\!\left(|\frac{\sqrt{2\tilde{D}} \sigma k}{\sigma_R}|\right) e^{ik\omega}$ 
%& $\langle \omega(t)\omega(t')\rangle \approx \delta(t-t')$ \\[0.8em]

\textbf{Model 3: Comoving OU} & & \\
Turning angle $\alpha $ 
\\  $\alpha = \xi_\alpha(t)$
& $P(\alpha) = \frac{\sqrt{2\tilde{D}\Delta t}}{2\pi} \sum_{k=-\infty}^{\infty} |k| K_1\!\left(|\frac{\sqrt{2\tilde{D}\Delta t} \sigma k}{\sigma_R}|\right) e^{ik\alpha}$ 
& $\langle \alpha(t)\alpha(t')\rangle \approx \delta(t-t')$ \\

Speed $s$ 
\\  $s = \tilde{\xi}_s(t)$
& $P(s)$ same as Polar OU 
& $\langle s(t)s(t')\rangle$ same as Polar OU \\[0.5em]

\end{tabular}
\end{ruledtabular}
\end{table*}

Overall, our comparative analysis confirms that all three models
consistently numerically reproduce the statistical properties of the
two-dimensional Cartesian OU process across all frames. As far as we
can tell, our analytical transformations between Cartesian, polar, and
comoving frames are exact, which is confirmed by the overall excellent
agreement of both probability distributions and autocorrelation
functions.

\section{Summary and conclusions}\label{sec:concl}

In summary, what we called the comoving frame is nothing else than the
coordinate frame suggested by Ross and Pearson to formulate a
two-dimensional random walk for modelling organismic movements
\cite{Ross04,Ross05,Pea06}. From these early origins comoving
coordinates propagated into ME in the form of the widely used
stochastic CRW model \cite{KaSh83,CPB08}. Turning angle and step
length distribution functions are in turn much extracted from
experimental data for moving organisms
\cite{TCB14,SRMVVBL20,Gunner24}. This theoretical framework is at
variance with stochastic models in active matter theory, which are all
formulated in the Cartesian coordinate frame, partially by using polar
coordinates \cite{RoSch11,GSR12,RBELS12,BeDiL16}. However, as we
pointed out in this article, the polar frame should not be confused
with the comoving frame. Defining stochastic dynamics in the latter
frame involves a non-trivial additional step beyond polar coordinates,
which is deriving a stochastic equation for the turning angle
dynamics, in addition to the speed dynamics. The important advantage
of the comoving frame is that it allows to formulate a set of
equations which more closely mimicks actual biological movements by
describing the fluctuations generating them in terms of equations that
are self-consistent in this frame. This approach matches to the
biophysical reasoning that self-propelled biological movements are
generated internally by an organism itself, and not by any external
noise in a fixed coordinate frame. Such a fixed Cartesian frame should
only be used to model the impact of external environmental conditions
on a moving organism. If a given stochastic process can be transformed
exactly between the Cartesian and the comoving frame, of course it
does not matter in which frame one expresses the given dynamics;
however, as we will argue further below, exact transformations might
not always be possible.

In our article, we have first defined the relevant three different
frames, with Cartesian and polar being trivial, as explained in
textbooks, but by identifying the comoving frame on this basis. The
essential idea of transforming stochastic processes between these
three different frames was first outlined for the simple random walk
as an example, and on this basis worked out in detail for the OU
process. While spherical and polar transformations of stochastic
dynamics and associated representations of OU probability
distributions are well known \cite{RoSch11,RBELS12,HoSt25}, to our
knowledge the autocorrelation functions for OU speed and orientation
angle have not been obtained before, which constitute first new
results. A more important finding is our, as far as we can tell, novel
turning angle distribution calculated exactly analytically for the OU
process, which does not seem to match to other famous circular
distributions like wrapped Gaussian or van Mises distributions.

Our most important result, however, is the self-consistent formulation
of the Cartesian OU process in the comoving frame. We argue that our
derivation is exact, as is in line with our numerical results. Interestingly, 
our comoving OU model generates a unimodal turning angle distribution, 
as is characteristic for the CRW of ME. However, on top of this
it features an exponentially decaying autocorrelation function for the
speed. To our knowledge,
such a generalised stochastic model defined in the comoving frame has
not yet been studied in ME, nor tested for organismic movements, even
though the importance of correlation functions to understand
organismic movements has been emphasized before \cite{LICCK12}.

Importantly, the comoving OU equations that we have derived form a
special case of the generalised Langevin dynamics in the comoving
frame stipulated {\em ad hoc} in Ref.~\cite{LCK13}, reading
\begin{eqnarray} 
  	 \dot{\alpha} &=& h(\alpha,s) + \xi_{\alpha;s}(t) \label{eq:ncm1}\\
  	 \dot{s}&=& g(\alpha,s) + \xi_{s;\alpha}(t)  \label{eq:ncm2}\: ,
\end{eqnarray}
where the forces on the right hand sides are split into coupled drift
terms and general noises.  Considering an overdamped version of
Eq.~\eqref{eq:ncm1} with a drift term that is linear in $\alpha$
reproduces the form derived for OU of Eq.~\eqref{eq:oucm} for the
turning angle \cite{LCK13}. The associated underdamped equation for
the speed clearly yields a special case of Eq.~\eqref{eq:ncm2}. Our
results for the comoving OU process thus demonstrate that the
theoretical framework heuristically provided by
Eqs.\eqref{eq:ncm1},\eqref{eq:ncm2} can be established, at least for
the special case of the OU process, from first principles. To our
knowledge, this is the first time that a comoving OU process has been
derived. Hence, our second main results is to have established, along
the lines of the OU process, a general theoretical framework that
clearly defines, and disentangles, the different coordinate frames for
generally transforming stochastic processes from the Cartesian into
the comoving frame. Our approach can thus be used to systematically
construct more general stochastic movement models than CRWs in the
comoving frame.

Especially, along these lines it would be interesting to merge the
approach put forward by active matter theory in terms of active
particle models with CRW models of ME \cite{KLK26}. Generalised
active particle models combining different types of stochastic
dynamics have already been considered recently \cite{DBG24}. In
forthcoming work we will show, by exploiting the conceptual framework
developed in this article, how to transform three generic active
particle models into the comoving frame. This will pave the way to
develop more general active particle models, and to make these ideas
attractive to applications in ME.

However, our present conjecture is that exact transformations between
the three different frames are only possible if the Cartesian
stochastic dynamics is Markovian, as for simple random walks and the
considered OU process. There is indeed a non-trivial Markovianisation
taking place viz.\ a loss of memory in the turning angle dynamics when
being wrapped onto the circle. It is indeed well known already that
there exist three different semi-Markovian L\'evy walk models, two
defined in the Cartesian frame and one in the comoving frame, which
cannot be transformed into each other \cite{ZFDB16}. This result
provides first evidence that it may in general not be possible to
transform non-Markovian processes defined in the Cartesian frame
exactly into the comoving frame. Indeed, for fractional Brownian
motion a comoving representation is, to our knowledge, not known
\cite{GKV25}.  However, there is a need to formulate representations
of more advanced stochastic processes in the comoving frame for
implementing self-consistent stochastic navigation in robots and
drones \cite{DOT16}.

Interestingly, if exact transformations of stochastic processes with
memory between these different frames are not possible, one might
conclude that active fluctuations generating self-propelled organismic
movements should {\em per se} not be defined in the Cartesian or polar
frames but exclusively self-consistently in the comoving frame. This
would, in turn, necessitate to develop a novel theory of stochastic
processes in the comoving frame, and to explore how such processes
then look like in the Cartesian frame. We finally remark that the
relevance of distinguishing between these different frames of
reference is already well-known to biologists studying organismic
movements. They call a world-centered frame, corresponding to the
Cartesian one, {\em allocentric} while the body-centered, comoving one
is known as {\em egocentric} \cite{Kla98}. In the former, animals
navigate according to fixed external landmarks while in the latter,
they integrate their path coordinates internally in the brain
\cite{Kla98,MoVl20}. These two frames are, in turn, represented in the
brain by different types of neurons, grid and place cells,
respectively \cite{EBB20}. There are thus plenty of reason to further
explore the theory of stochastic process in the comoving frame.

\begin{acknowledgments}
NLA wish to acknowledge to "Fundacion Politecnico" and the School of
Mathematical Sciences from Queen Mary University of London for the
financial support of this research. NLA and RK thank Alessia Gentili,
Giorgio Volpe, Adrian Baule and Wolfram Just for very helpful discussions.
\end{acknowledgments}

\appendix

%\section{Appendixes}

\section{Autocorrelations in Speed and Orientation Random Walk}
\label{app:no_autocorr}

We consider a two--dimensional random walk defined by
\begin{equation}
\mathbf{x}_{n+1}=\mathbf{x}_n+\Delta t\,\mathbf{v}_n .
\label{eq:rw}
\end{equation}

Since the velocities $\mathbf{v}_n$ are i.i.d.\ for different time
indices, and since $(s_n,\beta_n)$ are deterministic functions of
$\mathbf{v}_n$, it follows that
\begin{equation}
(s_n,\beta_n)\;\perp\!\!\!\perp\;(s_{n+k},\beta_{n+k}),
\qquad k\neq 0 .
\label{eq:temporal_indep}
\end{equation}

\subsection{Autocorrelation of the speed}

The autocovariance of the speed is defined as
\begin{equation}
  C_s(n,k)=\langle s_k s_{k+n}\rangle-\langle s_k\rangle^2 ,
\label{eq:Cs}
\end{equation}
where the angular brackets denote an ensemble average.  For $n\neq 0$,
temporal independence \eqref{eq:temporal_indep} implies
\[
\langle s_k s_{k+n}\rangle
=\langle s_k\rangle\langle s_{k+n}\rangle .
\]
By stationarity, $\langle s_k\rangle=\langle s_{k+n}\rangle=\langle s\rangle$,
and therefore
\begin{equation}
C_s(n,k)=0,
\qquad n\neq 0 .
\end{equation}

\subsection{Autocorrelation of the orientation angle}

The autocovariance of the orientation angle is defined as
\begin{equation}
C_\beta(n,k)=\langle \beta_k \beta_{k+n}\rangle-\langle \beta_k\rangle^2 .
\label{eq:Cbeta}
\end{equation}
For $n\neq 0$, Eq.~\eqref{eq:temporal_indep} yields
\[
\langle \beta_k \beta_{k+n}\rangle
=\langle \beta_k\rangle\langle \beta_{k+n}\rangle .
\]
Since $\beta$ is uniformly distributed in $[0,2\pi)$,
\begin{equation}
\langle \beta\rangle=\frac{1}{2\pi}\int_0^{2\pi}\beta\,d\beta=\pi .
\end{equation}
Consequently,
\begin{equation}
C_\beta(n,k)=0,
\qquad n\neq 0 .
\end{equation}

\subsection{Autocorrelation of the turning angle}

We define the turning angle as the difference between two successive orientation angles,
\begin{equation}
\alpha_n=\beta_{n+1}-\beta_n 
\label{eq:turning}
\end{equation}
with $\alpha_n\in (-\pi,\pi]$. Because $\beta_{n+1}$ and $\beta_n$ are
independent and uniformly distributed, the turning angle $\alpha_n$ is
itself uniformly distributed on $(-\pi,\pi]$.  Moreover, for $n\neq
0$, the pairs $(\beta_{k+1},\beta_k)$ and
$(\beta_{n+k+1},\beta_{n+k})$ are independent.  Therefore,
\begin{equation}
\langle \alpha_k \alpha_{k+n}\rangle
=\langle \alpha_k\rangle\langle \alpha_{k+n}\rangle ,
\qquad n\neq 0 .
\end{equation}

The autocovariance of the turning angle,
\begin{equation}
C_\alpha(n,k)
=\langle \alpha_k \alpha_{k+n}\rangle-\langle \alpha_k\rangle^2 ,
\label{eq:Calpha}
\end{equation}
thus vanishes for all nonzero lags,
\begin{equation}
C_\alpha(n,k)=0,
\qquad n\neq 0 .
\end{equation}

\section{It\^o transformation} \label{AITO}

\newcommand{\D}{\tilde{D}}
\newcommand{\vv}{\mathbf{v}}
\newcommand{\W}{\mathbf{W}}
\newcommand{\Wx}{W_x}
\newcommand{\Wy}{W_y}
\newcommand{\WS}{W_S}
\newcommand{\Wbeta}{W_\beta}
\newcommand{\xiV}{\boldsymbol{\xi}_{\mathbf{v}}}

We start with the two-dimensional stochastic differential equation for the velocity $\vv = (v_x, v_y)$:
\begin{equation}
d\vv = -\frac{\gamma}{m} \vv  dt + \sqrt{2\D}  d\W(t),
\label{sde_original}
\end{equation}
where $d\W(t) = (d\Wx, d\Wy)$ is a vector of independent Wiener processes.

We define the polar coordinates:
\begin{align*}
s &= \sqrt{v_x^2 + v_y^2}, \\
\beta &= \arctan\left(\frac{v_y}{v_x}\right).
\end{align*}

We apply It\^o's lemma to find the stochastic differential equations for $s$ and $\beta$.
Let $s = f(v_x, v_y) = (v_x^2 + v_y^2)^{1/2}$. The partial derivatives are:
\begin{align*}
\frac{\partial s}{\partial v_x} &= \frac{v_x}{s}, \quad
\frac{\partial s}{\partial v_y} = \frac{v_y}{s}, \\
\frac{\partial^2 s}{\partial v_x^2} &= \frac{1}{s} - \frac{v_x^2}{s^3}, \quad
\frac{\partial^2 s}{\partial v_y^2} = \frac{1}{s} - \frac{v_y^2}{s^3}.
\end{align*}

The drift and diffusion coefficients for $v_x$ and $v_y$ are:
\[
\mu_x = -\frac{\gamma}{m} v_x, \quad \mu_y = -\frac{\gamma}{m} v_y, \quad \sigma_x = \sqrt{2\D}, \quad \sigma_y = \sqrt{2\D}.
\]

By It\^o's lemma:
\begin{align*}
ds &= \left[ \mu_x \frac{\partial s}{\partial v_x} + \mu_y \frac{\partial s}{\partial v_y} 
+ \frac{1}{2} \sigma_x^2 \frac{\partial^2 s}{\partial v_x^2} + \frac{1}{2} \sigma_y^2 \frac{\partial^2 s}{\partial v_y^2} \right] dt \\
&\quad + \sigma_x \frac{\partial s}{\partial v_x} d\Wx + \sigma_y \frac{\partial s}{\partial v_y} d\Wy.
\end{align*}

Substituting the derivatives:
\begin{align*}
ds &= \Bigg[ \left(-\frac{\gamma}{m} v_x\right) \frac{v_x}{s} + \left(-\frac{\gamma}{m} v_y\right) \frac{v_y}{s} \\
&\quad + \frac{1}{2} (2\D) \left( \frac{1}{s} - \frac{v_x^2}{s^3} \right) + \frac{1}{2} (2\D) \left( \frac{1}{s} - \frac{v_y^2}{s^3} \right) \Bigg] dt \\
&\quad + \sqrt{2\D} \frac{v_x}{s} d\Wx + \sqrt{2\D} \frac{v_y}{s} d\Wy.
\end{align*}

Simplifying the drift term:
\begin{align*}
& \left[ -\frac{\gamma}{m} \frac{v_x^2 + v_y^2}{s} + \D \left( \frac{1}{s} - \frac{v_x^2}{s^3} \right) + \D \left( \frac{1}{s} - \frac{v_y^2}{s^3} \right) \right] dt \\
&= \left[ -\frac{\gamma}{m} s + \D \left( \frac{2}{s} - \frac{v_x^2 + v_y^2}{s^3} \right) \right] dt \\
&= \left[ -\frac{\gamma}{m} s + \D \left( \frac{2}{s} - \frac{s^2}{s^3} \right) \right] dt \\
&= \left[ -\frac{\gamma}{m} s + \D \left( \frac{2}{s} - \frac{1}{s} \right) \right] dt \\
&= \left[ -\frac{\gamma}{m} s + \frac{\D}{s} \right] dt.
\end{align*}

The diffusion term is:
\[
\sqrt{2\D} \frac{v_x}{s} d\Wx + \sqrt{2\D} \frac{v_y}{s} d\Wy.
\]
This can be written as $\sqrt{2\D}  d\WS$, where $d\W_s = \frac{v_x}{s} d\Wx + \frac{v_y}{s} d\Wy$ is a Wiener process.

Thus, the stochastic differential equation for $S$ is:
\begin{equation}
ds = \left( -\frac{\gamma}{m} s + \frac{\D}{s} \right) dt + \sqrt{2\D}  d\W_s.
\label{sde_S}
\end{equation}

Let $\beta = f(v_x, v_y) = \arctan\left(\frac{v_y}{v_x}\right)$. The partial derivatives are:
\begin{align*}
\frac{\partial \beta}{\partial v_x} &= -\frac{v_y}{v_x^2 + v_y^2} = -\frac{v_y}{s^2}, \\
\frac{\partial \beta}{\partial v_y} &= \frac{v_x}{v_x^2 + v_y^2} = \frac{v_x}{s^2}, \\
\frac{\partial^2 \beta}{\partial v_x^2} &= \frac{2 v_x v_y}{s^4}, \\
\frac{\partial^2 \beta}{\partial v_y^2} &= -\frac{2 v_x v_y}{s^4}, \\
\frac{\partial^2 \beta}{\partial v_x \partial v_y} &= \frac{v_y^2 - v_x^2}{s^4}.
\end{align*}

By It\^o's lemma:
\begin{align*}
d\beta &= \Bigg[ \mu_x \frac{\partial \beta}{\partial v_x} + \mu_y \frac{\partial \beta}{\partial v_y} 
+ \frac{1}{2} \sigma_x^2 \frac{\partial^2 \beta}{\partial v_x^2} + \frac{1}{2} \sigma_y^2 \frac{\partial^2 \beta}{\partial v_y^2} \\
&\quad + \frac{1}{2} \sigma_x \sigma_y \frac{\partial^2 \beta}{\partial v_x \partial v_y} \Bigg] dt \\
&\quad + \sigma_x \frac{\partial \beta}{\partial v_x} d\Wx + \sigma_y \frac{\partial \beta}{\partial v_y} d\Wy.
\end{align*}

Since $\sigma_x = \sigma_y = \sqrt{2\D}$ and the noises are independent, the cross term is zero. Substituting:
\begin{align*}
d\beta &= \Bigg[ \left(-\frac{\gamma}{m} v_x\right) \left(-\frac{v_y}{s^2}\right) + \left(-\frac{\gamma}{m} v_y\right) \left(\frac{v_x}{s^2}\right) \\
&\quad + \frac{1}{2} (2\D) \frac{2 v_x v_y}{s^4} + \frac{1}{2} (2\D) \left(-\frac{2 v_x v_y}{s^4}\right) \Bigg] dt \\
&\quad + \sqrt{2\D} \left(-\frac{v_y}{s^2}\right) d\Wx + \sqrt{2\D} \left(\frac{v_x}{s^2}\right) d\Wy.
\end{align*}

The drift terms cancel:
\[
\left(-\frac{\gamma}{m} v_x\right) \left(-\frac{v_y}{s^2}\right) + \left(-\frac{\gamma}{m} v_y\right) \left(\frac{v_x}{s^2}\right) = \frac{\gamma}{m} \frac{v_x v_y}{s^2} - \frac{\gamma}{m} \frac{v_x v_y}{s^2} = 0,
\]
and
\[
\frac{1}{2} (2\D) \frac{2 v_x v_y}{s^4} + \frac{1}{2} (2\D) \left(-\frac{2 v_x v_y}{s^4}\right) = \D \frac{2 v_x v_y}{s^4} - \D \frac{2 v_x v_y}{s^4} = 0.
\]
Thus, the drift is zero.

The diffusion term is:
\[
\sqrt{2\D} \left( -\frac{v_y}{s^2} d\Wx + \frac{v_x}{s^2} d\Wy \right).
\]
This can be written as $\frac{\sqrt{2\D}}{s} d\Wbeta$, where $d\Wbeta = -\frac{v_y}{s} d\Wx + \frac{v_x}{s} d\Wy$ is a Wiener process.

Thus, the stochastic differential equation for $\beta$ is:
\begin{equation}
d\beta = \frac{\sqrt{2\D}}{s}  d\Wbeta.
\label{sde_beta}
\end{equation}

In summary, the transformed stochastic differential equations in polar coordinates are:
\begin{align}
ds &= \left( -\frac{\gamma}{m} s + \frac{\D}{s} \right) dt + \sqrt{2\D}  d\W_s, \label{sde_S_final} \\
d\beta &= \frac{\sqrt{2\D}}{s}  d\Wbeta, \label{sde_beta_final}
\end{align}
where $d\WS$ and $d\Wbeta$ are independent Wiener processes.

\section{Autocorrelation Function for the Speed in th OU} \label{acs}
To obtain the mean value of the speed, we examine the steady-state behavior of Eq.(\ref{eq:itos}). In this limit, the left-hand side becomes zero:
\begin{equation}
0=-\frac{\gamma}{m} s_0+\frac{\tilde{D}}{s_0}\:,
\end{equation}
where  $s_0$ represents the mean speed in the steady-state. Solving for $s_0$ we find:
\begin{equation}
s_0^2=\frac{\tilde{D}m}{\gamma}
\end{equation}
which implies that $s_0=\sqrt{\frac{\tilde{D}m}{\gamma}}$.

Secondly, to calculate the autocorrelations for $s$, we introduce a linear expansion around $s_0$ for $s(t)$:
\begin{equation}
s(t)=s_0 + \delta s(t)
\end{equation}
where $\delta s(t)$ represents the fluctuations around the steady-state value $s_0$. Substituting this into Eq.(\ref{eq:itos}):
\begin{equation}
\frac{d\left(s_0 +\delta s(t)\right)}{dt}=-\frac{\gamma}{m}\left(s_0 +\delta s(t)\right) +\frac{\tilde{D}}{s_0 +\delta s(t)} +\sqrt{2\tilde{D}}\xi_{s}(t) \label{sdescv}
\end{equation}
From this, we observe that the term $\frac{\tilde{D}}{s_0 +\delta s(t)}$ can be rewritten as:
\begin{equation}
\frac{\tilde{D}}{s_0 +\delta s(t)}=\frac{\tilde{D}}{s_0}\frac{1}{1+\frac{\delta s(t)}{s_0}}
\end{equation}
Here, we can expand the factor $\frac{1}{1+\frac{\delta s(t)}{s_0}}$ in a Taylor series for small values of $\frac{\delta s(t)}{s_0}$:
\begin{equation}
\frac{\tilde{D}}{s_0}\frac{1}{1+\frac{\delta s(t)}{s_0}} \approx \frac{\tilde{D}}{s_0} - \frac{\tilde{D}}{s_0^2}\delta s(t)
\end{equation}
Thus, Eq.(\ref{sdescv}) becomes:
\begin{equation}
\frac{d\left(s_0 +\delta s(t)\right)}{dt}=-\frac{\gamma}{m}\left(s_0 +\delta s(t)\right) +\frac{\tilde{D}}{s_0} - \frac{\tilde{D}}{s_0^2}\delta s(t) +\sqrt{2\tilde{D}}\xi_{s}(t) .\label{sdescvm}
\end{equation}
Applying the linearity of the differential operator on the left side and substituting $s_0=\sqrt{\frac{\tilde{D}m}{\gamma}}$, several terms cancel out:
\begin{equation}
\frac{d\delta s(t)}{dt}=-\frac{\gamma}{m}\delta s(t) - \frac{\tilde{D}}{s_0^2}\delta s(t) +\sqrt{2\tilde{D}}\xi_{s}(t) \label{sdescv2}
\end{equation}
Simplifying, we get:
\begin{equation}
\frac{d\delta s(t)}{dt}=-2\frac{\gamma}{m}\delta s(t) +\sqrt{2\tilde{D}}\xi_{s}(t) \label{sdescv3}
\end{equation}
We notice that Eq.(\ref{sdescv3}) has the same form as an OU process. Hence, its autocorrelation can be calculated similarly as:
\begin{equation}
\langle \delta s(t)\delta s(t')\rangle=\frac{\tilde{D}m}{2\gamma} e^{-2\frac{\gamma}{m}\left|t-t'\right|}.
\end{equation}

Now, for our original variable $s(t)$, the autocorrelation is:
\begin{equation}
\langle s(t)s(t')\rangle =\langle s_0\rangle^2 + \langle \delta s(t)\delta s(t')\rangle
\end{equation}
Finally, we can compute the autocorrelation function:
\begin{equation}
\langle s(t)s(t')\rangle -\langle s_0\rangle^2= \frac{\tilde{D}m}{2\gamma} e^{-2\frac{\gamma}{m}\left|t-t'\right|} \label{AC-Speed}
\end{equation}

\section{Autocorrelation of $\cos \beta$ in OU} \label{AP-beta}

Consider the orientation angle $\beta(t)$ evolving according to the polar OU process:
\begin{equation}
\frac{d\beta}{dt} = \frac{\sqrt{2 \tilde{D}}}{s} \, \xi_\beta(t),
\end{equation}
where $\xi_\beta(t)$ is Gaussian white noise with zero mean and unit variance. For a discrete time increment $\Delta t$, the angular increment can be approximated as
\begin{equation}
\Delta \beta = \beta(t+\Delta t) - \beta(t) \sim \mathcal{N}\Big(0, \sigma_\beta^2\Big),
\end{equation}
with variance
\begin{equation}
\sigma_\beta^2 = \frac{2 \tilde{D} \Delta t}{\langle s^2 \rangle}.\label{varcosba}
\end{equation}

With the assumption of $\Delta \beta(t)$ as Gaussian distributed, the autocorrelation of $\cos \beta$ is then
\begin{align}
\langle \cos \beta(t) \cos \beta(t+\tau) \rangle 
&= \frac{1}{2} \left\langle e^{i (\beta(t) - \beta(t+\tau))} + e^{i (\beta(t) + \beta(t+\tau))} \right\rangle \nonumber \\
&\approx \frac{1}{2} \exp\Big[-\frac{1}{2} \mathrm{Var}(\beta(t+\tau)-\beta(t)) \Big] ,\label{approx_beta_result}
\end{align}

Combining with~\eqref{varcosba} gives
\begin{equation}
\langle \cos\beta(t)\cos\beta(t+\tau)\rangle
\approx
\frac{1}{2}\,
\exp\!\Big[
-\tilde D\,\tau\,
\Big\langle s^{-2}\Big\rangle
\Big],
\label{eq:cosbeta_corr}
\end{equation}
where $\langle s^{-2}\rangle$ can be approximated by numerical fitting to the simulations of the models.
This expression shows that the autocorrelation of $\cos\beta$ decays
exponentially, with a rate determined by both the OU relaxation time
and the variance of the angular increments $\sigma_\beta^2$.

\section{Autocorrelation of $\omega(t)$ and $\alpha$ in OU}\label{acdotbeta}
The autocorrelation of $\omega(t)$ is defined as:
\begin{equation}
\langle \omega(t) \omega(t') \rangle = \left\langle \frac{\sqrt{2\tilde{D}}}{s(t)} \xi_\beta(t) \cdot \frac{\sqrt{2\tilde{D}}}{s(t')} \xi_\beta(t') \right\rangle.
\end{equation}
Using the property of white noise $\langle \xi_\beta(t) \xi_\beta(t') \rangle = \delta(t - t')$, we simplify the expression:
\begin{equation}
\langle \omega(t) \omega(t') \rangle = \frac{2\tilde{D}}{\langle s(t) s(t')\rangle} \delta(t - t').
\end{equation}
which shows that the angular velocity is simply a delta correlated variable. Accordingly, for the turning angle, which is directly related to the angular velocity via $\alpha=\omega \sqrt{dt}$, we obtain that the autocorrelations are
\begin{equation}
    \langle \alpha(t) \alpha(t') \rangle\approx\delta(t-t')    
\end{equation}

\section{Calculation of the Probability Distribution for the Turning Angle in the OU}\label{ap:tapdf}

According to the Itô equation (\ref{eq:itop}), the instantaneous angular velocity $\dot{\beta}$ is given by the term $\sqrt{2\tilde{D}}\frac{\xi_{\beta}(t)}{s(t)}$. To calculate the probability distribution of the turning angle $\alpha$, we note that $\alpha/\Delta t = \dot{\beta}$. Therefore, the scale parameter of the resulting probability density function (PDF) will be modified by the time step $\Delta t$. The steps to find such a PDF are as follows:

First, we perform a change of variables for the term $\sqrt{2\tilde{D}}\frac{\xi_{\beta}(t)}{s(t)}$
\begin{align}
Z&=a\frac{X}{Y} \\
W&=Y
\end{align}
where $X$ represents a Gaussian white noise with standard deviation $\sigma$, $Y$ represents an exponentially correlated Rayleigh distribution with scale parameter $\sigma_R$, and $a$ is a constant. The joint distribution $P(z,w)$ can be found using the Jacobian of the transformation:
\begin{equation}
P(z,w)=P(x)P(y)J
\end{equation}
The Jacobian $J$ is calculated as:
\[J=\det\begin{pmatrix}
\frac{\partial x}{\partial z} & \frac{\partial x}{\partial w} \\
\frac{\partial y}{\partial z} & \frac{\partial y}{\partial w}
\end{pmatrix}=\frac{w}{a}\]
Now, we can write the joint distribution as:
\begin{equation}
P(z,w)=\frac{1}{\sqrt{2 \pi \sigma^2}} \exp\left( -\frac{\left(\frac{zw}{a}\right)^2}{2 \sigma^2} \right) \frac{w}{\sigma_R^2} \exp\left( -\frac{w^2}{2 \sigma_R^2} \right) \frac{w}{a}
\end{equation}
which can be simplified to:
\begin{equation}
P(z,w)=\frac{1}{\sqrt{2 \pi \sigma^2}} \frac{w^2}{a\sigma_R^2} \exp\left( -\frac{w^2}{2}\left( \frac{z^2}{a^2\sigma^2}+\frac{1}{\sigma_R ^2}\right) \right)
\end{equation}

Since we are interested in the marginal distribution for $z$, we integrate the joint distribution with respect to $w$ from $0$ to $\infty$:
\begin{equation}
P(z)=\int_{0}^{\infty} \frac{1}{\sqrt{2 \pi \sigma^2}} \frac{w^2}{a\sigma_R^2} \exp\left( -\frac{w^2}{2}\left( \frac{z^2}{a^2\sigma^2}+\frac{1}{\sigma_R ^2}\right) \right) dw
\end{equation}
Upon integration, we find that:
\begin{equation}
P(z)=\frac{1}{2 a\sigma\sigma_R^2 \left( \frac{1}{\sigma_R^2}+\frac{z^2}{a^2\sigma^2 }\right)^{3/2}}
\end{equation}
Simplifying this expression yields:
\begin{equation}
P(z)=\frac{a^2\sigma^2}{2\sigma_R^2 \left( \frac{a^2 \sigma^2}{\sigma_R^2}+z^2\right)^{3/2}} \label{pz}
\end{equation}
This is the PDF for the variable on the real line in the comoving frame. However, for describing particle turning angles, a representation within the range $[-\pi,\pi]$ is more natural. To achieve this, we wrap our PDF to the circle by applying the Poisson Summation Formula:
\begin{equation}
P(\theta) = \frac{1}{2\pi} \sum_{k=-\infty}^{\infty} \hat{P}(k) e^{ik\theta}
\end{equation}
where $\hat{P}(k)$ is the Fourier transform (FT) of $P_Z(z)$ from Eq.(\ref{pz}). We calculate the FT:
\begin{equation}
\hat{P}(k) = \int_{-\infty}^{\infty} \frac{ a^2\sigma^2}{2\sigma_R^2 \left( \frac{a^2\sigma^2}{\sigma_R^2}+\theta^2\right)^{3/2}} e^{-ik\theta} d\theta
\end{equation}
This integral evaluates to:
\begin{equation}
\hat{P}(k) =a \left|\frac{\sigma}{\sigma_R} k\right| K_1\left(\left|\frac{a\sigma k}{\sigma_R}\right|\right)
\end{equation}
where $K_1(z)$ is the modified Bessel function of the second kind of order $1$.
Now, substituting this into the Poisson formula:
\begin{equation}
P(\theta) = \frac{a}{2\pi} \sum_{k=-\infty}^{\infty} \left|\frac{\sigma}{\sigma_R} k\right| K_1\left(\left|\frac{a\sigma k}{\sigma_R}\right|\right) e^{i k \theta}
\end{equation}
This represents the PDF for the ratio of a Gaussian distribution and a Rayleigh distribution in the comoving frame, within the range $[-\pi,\pi]$.

Substituting $a=\sqrt{2\tilde{D}}$ to find the angular velocity PDF:
\begin{equation}
P(\omega) = \frac{\sqrt{2\tilde{D}}}{2\pi} \sum_{k=-\infty}^{\infty} \left|\frac{\sigma}{\sigma_R} k\right| K_1\left(\left|\frac{\sqrt{2\tilde{D}} \sigma k}{\sigma_R}\right|\right) e^{i k \omega} \label{pdf-angularv}
\end{equation}
Here, the variable $\omega$ is in the interval $[-\pi,\pi]$ with units of radians per second (rad/s), as this is an angular velocity.

In the case of the PDF of the turning angle $\alpha$, we need to substitute $a=\sqrt{2\tilde{D}\Delta t}$, obtaining:
\begin{equation}
P(\alpha,\Delta t) = \frac{\sqrt{2\tilde{D}\Delta t}}{2\pi} \sum_{k=-\infty}^{\infty} \left|\frac{\sigma}{\sigma_R} k\right| K_1\left(\left|\frac{\sqrt{2\tilde{D}\Delta t} \sigma k}{\sigma_R}\right|\right) e^{i k \alpha}
\end{equation}
Here, we observe a dependence on the time step $\Delta t$ chosen for the simulations, as this refers to the turning angle, which will have a different distribution depending on how frequently the dynamics are sampled. For this case, the interval is between $[-\pi,\pi]$ with units of radians.\\

\bibliography{summ48}% Produces the bibliography via BibTeX.

\end{document}